\documentclass[aps,pra,showpacs,superscriptaddress,twocolumn,final]{revtex4-1}
\usepackage{amsmath,amsfonts,amsthm}
\usepackage{graphicx}

\newtheorem{proposition}{Proposition}

\DeclareMathOperator{\tr}{tr}

\newcommand{\bra}[1]{{\langle{#1}\rvert}}
\newcommand{\ket}[1]{{\lvert{#1}\rangle}}
\newcommand{\ketbra}[2]{{\ket{#1}\!\bra{#2}}}

\newcommand{\proj}[1]{\ketbra{#1}{#1}}
\newcommand{\id}{\openone}
\newcommand{\Hilbert}{{\mathcal H}}
\newcommand{\complexes}{{\mathbb C}}

\newcommand{\idop}{{\mathfrak I}}
\newcommand{\Xambda}{{\mathfrak E}}

\begin{document}

\title{Asymptotically perfect discrimination in the LOCC paradigm}

\author{M.~Kleinmann}
\affiliation{Department Physik,
             Universit\"at Siegen,
             Walter-Flex-Str.~3,
             D-57068 Siegen, Germany}
\affiliation{Institut f\"ur Quantenoptik und Quanteninformation,
             \"Osterreichische Akademie der Wissenschaften,
             Technikerstr.~21A,
             A-6020 Innsbruck, Austria}

\author{H.~Kampermann}

\author{D.~Bru\ss}
\affiliation{Institut f\"ur Theoretische Physik III, 
             Heinrich-Heine-Universit\"at D\"usseldorf,
             D-40225 D\"usseldorf, Germany}

\begin{abstract}
We revisit the problem of discriminating orthogonal quantum states within the 
 local quantum operation and classical communication (LOCC) paradigm.
Our particular focus is on the asymptotic situation where the parties have 
 infinite resources and the protocol may become arbitrarily long.
Our main result is a necessary condition for perfect asymptotic LOCC 
 discrimination.
As an application, we prove that for complete product bases, unlimited 
 resources are of no advantage.
On the other hand, we identify an example, for which it still remains undecided 
 whether unlimited resources are superior.
\end{abstract}


\maketitle

\section{Introduction}
An important concept in quantum information theory is the paradigm known as 
 ``local operations and classical communication'' (LOCC).
It specifies the operational power of two or more parties which only have local 
 access to a distributed quantum system but are equipped with a classical 
 communication channel.
A typical question now is, whether a certain task that usually is trivial to 
 perform with global access can be accomplished within this restricted set of 
 operations.
Prominent such examples are entanglement distillation, entanglement 
 transformations, or local state discrimination, and results from such examples 
 have strong influence on central topics in quantum information theory, e.g.\ 
 in entanglement classification and quantification or in quantum communication 
 theory \cite{Horodecki:2009RMP, Guhne:2009PR}.

Here, we will focus on the local discrimination of orthogonal states, i.e., 
 states, which can be discriminated perfectly by a global measurement.
This situation has been studied extensively in the literature, cf.\ e.g.\ 
 Ref.~\cite{Ghosh:2001PRL, Terhal:2001PRL, Horodecki:2003PRL, Gosh:2004PRA, 
 Chen:2001PRA, Chen:2003PRA, *Chen:2004PRA,  Nathanson:2005:JMP, 
 Watrous:2005PRL, Duan:2010PRA}, and some of the results are quite 
 counter-intuitive.
For example, it is always possible to perfectly discriminate two arbitrary 
 orthogonal states \cite{Walgate:2000PRL}, while there exist product bases 
 which cannot be discriminated perfectly by means of LOCC 
 \cite{Bennett:1999PRA}.

An LOCC discrimination protocol in general consists of several rounds, where in 
 each round one party performs a measurement and communicates the results to 
 all parties.
Due to the existence of ``weak measurements'' \cite{Renninger:1960ZFP, 
 Dicke:1981AJP} it is not clear that perfect discrimination can be achieved in 
 a finite number of such rounds.
From a physical point of view, the question of perfect distinguishability is 
 not particularly meaningful, since unavoidable experimental imperfections will 
 always impede perfect measurement results.
Rather it would be interesting to know, whether with increasing experimental 
 effort, one can get arbitrarily close to perfect discrimination.
This asymptotic case has already been noticed and approached in 
 Ref.~\cite{Bennett:1999PRA}, but to our knowledge only in 
 Ref.~\cite{Rinaldis:2004PRA} this question has been considered again, while 
 the majority of the work on LOCC discrimination explicitly is limited to 
 perfect discrimination in a finite number of rounds (cf.\ e.g.\ 
 Ref.~\cite{Chen:2001PRA, Chen:2003PRA, *Chen:2004PRA, Nathanson:2005:JMP, 
 Watrous:2005PRL, Duan:2010PRA}) or to the more general class of stochastic 
 LOCC measurements (or separable measurements \footnote{%
A stochastic LOCC measurement is a measurement that can be implemented by means 
 of LOCC with a certain probability $p>0$, while with probability $1-p$ the 
 measurement will fail.
Stochastic LOCC measurements are exactly those with separable effects as in 
 Eq.~\eqref{e19171} and hence the alternative name \emph{separable 
 measurements} has also been used.%
}), cf.\ e.g.\ Ref.~\cite{Ghosh:2001PRL, Terhal:2001PRL, Horodecki:2003PRL, 
 Gosh:2004PRA}.
So far it is actually unclear whether the asymptotic consideration may yield a 
 different result than the finite analysis.

In this contribution, we now revisit the problem of perfect discrimination by 
 asymptotic LOCC.
Our main result is a general necessary condition for such a discrimination to 
 be possible, cf.\ Proposition~\ref{p25379}.
The proof of this result uses a variant of the protocol splitting technique 
 introduced in Ref.~\cite{Bennett:1999PRA}.
We, however, do not rely on a continuous measurement process, but rather show 
 that a finite enlargement of the protocol suffices in order to employ the 
 protocol splitting.
As an application of Proposition~\ref{p25379}, we show that a product basis can 
 be discriminated asymptotically if and only if it can be discriminated by 
 finite means.
This also gives an analytical proof of the numerical findings in 
 Ref.~\cite{Bennett:1999PRA}.
(A similar result regarding unextendible product bases was stated in 
 Ref.~\cite{Rinaldis:2004PRA}, however we question the validity of this proof, 
 cf.\ our Remark below Proposition~\ref{p23315}.)
Finally, we study an example provided by Duan \emph{et al.} 
 \cite{Duan:2009IIT}, for which it is known that it cannot be discriminated by 
 any finite protocol, while it can be discriminated perfectly by stochastic 
 LOCC.
For this example, using our result, we cannot exclude that asymptotic LOCC 
 could achieve perfect discrimination.

Our paper is organized as follows:
In Sec.~\ref{s1693} we thoroughly define our notion of asymptotic LOCC 
 discrimination and analyze possible generalizations.
In Sec.~\ref{s20497} we prove our main result, which is summarized in 
 Proposition~\ref{p25379}.
We then discuss two examples in Sec.~\ref{s17873} before we conclude in 
 Sec.~\ref{s25406}.

\section{Asymptotic LOCC discrimination}\label{s1693}
In our scenario we aim at discriminating a certain family of multipartite mixed 
 states $(\rho_\mu)$, where $\rho_\mu$ are density operators on a 
 finite-dimensional Hilbert space $\Hilbert= \bigotimes_r \Hilbert^{(r)}$.
We will first define a general notion of finite LOCC measurements and then 
 describe the transition from those finite measurements to the asymptotic 
 situation.

\subsection{Finite LOCC measurements}
The most general quantum measurement with $n$ outcomes is described by a 
 positive operator valued measure (POVM), i.e., a finite family $(E_k)$ of $n$ 
 positive semi-definite operators (or \emph{effects}) on $\Hilbert$ obeying 
 $\sum_k E_k= \id$.
The probability to obtain the outcome $k$ for a state $\rho_\mu$ is then given 
 by $\tr(\rho_\mu E_k)$.
Hence a measurement can be written as the mapping $\Xambda\colon X \mapsto 
 (\tr[X E_k])$ from the set of operators into $\complexes^n$, where $0\le 
 \Xambda(\rho)_k\le 1$ for any state~$\rho$.

Any POVM can be implemented by a physical measurement device and vice versa, 
 any such device corresponds to a unique POVM.
If the physical setup is limited to the LOCC paradigm then each effect $E_k$ 
 will be a sum of positive semi-definite product operators 
 \cite{Barnum:1998PRA, Bennett:1999PRA},
\begin{equation}\label{e19171}
 E_k = \sum_j \bigotimes_r E_{k,j}^{(r)} \text{ with }
 E_{k,j}^{(r)}\ge 0.
\end{equation}
However, as first shown by Bennett \emph{et al.} in 
 Ref.~\cite{Bennett:1999PRA}, the converse statement does not hold in general.

We call a measurement a \emph{finite LOCC measurement}, if it can be 
 implemented by an LOCC protocol, using only finite dimensional ancilla 
 systems, measurements with a finite number of outcomes and which is guaranteed 
 to terminate after a certain number of rounds.
The intuition behind this restriction is a realistic experimental setup, where 
 the effective dimension of the Hilbert space shall be finite, the classical 
 communication channel has limited capacity, and the experiment cannot be kept 
 stable for an infinite time span.

\subsection{Deviation from perfect discrimination}
For our goal of perfect discrimination of orthogonal states, we now measure the 
 deviation from perfect discrimination $d(\Xambda)$ for an arbitrary 
 measurement $\Xambda$.
Therefore we assume that for some fixed set of states $(\rho_\mu)$, 
 $d(\Xambda)$ is a non-negative real number such that $d(\Xambda)= 0$ implies 
 that $\Xambda$ achieves perfect discrimination of $(\rho_\mu)$.
Then we define the \emph{asymptotic deviation} $\hat d$ as the infimum of $d$ 
 over all finite LOCC measurements.
In particular, if $\hat d= 0$ then for any deviation $\varepsilon>0$ we can 
 find a finite LOCC measurement $\Xambda^\varepsilon$, such that 
 $d(\Xambda^\varepsilon)< \varepsilon$.

The deviation measure has to be chosen carefully, as a trivial (but meaningful) 
 choice for the deviation is e.g.\ the measure $d_\mathrm{finite}$, which 
 yields $1$ whenever the measurement fails to achieve perfect discrimination 
 and $0$ in the case of perfect discrimination.
Then $\hat d_\mathrm{finite}= 0$ if and only if there exists a finite LOCC 
 measurement that achieves perfect discrimination.

Typically we would be rather interested, whether e.g.\ the mean failure 
 probability could approach zero as the LOCC measurement becomes more and more 
 expensive.
We thus define the deviation measure $d_\mathrm{mf}(\Xambda)$ to be the minimal 
 mean failure probability over any possible classical post-processing of 
 $\Xambda$, i.e.,
\begin{equation}\label{e3208}
 d_\mathrm{mf}(\Xambda) = 1- \sum_k \max_\mu (p_\mu \Xambda(\rho_\mu)_k),
\end{equation}
 with some arbitrary \emph{a priori} probabilities $p_\mu> 0$ obeying $\sum_\mu 
 p_\mu= 1$.
(The interpretation of this measure is as follows:
Assume that the state $\rho_\mu$ is prepared with probability $p_\mu$ and we 
 use the measurement $\Xambda$ in order to learn about the index $\mu$.
Given the measurement result $k$, the strategy which minimizes the probability 
 of a failure is the one in which  we announce the index $\mu$ maximizing 
 $p_\mu \Xambda(\rho_\mu)_k$.)

In Ref.~\cite{Bennett:1999PRA}, in contrast, an entropy based measure was used 
 for the deviation measure, namely the conditional entropy
\begin{equation}\label{e21158}
 d_\mathrm{ce}(\Xambda)= H(S|K)\equiv H(S,K) - H(K)
\end{equation}
 where $S$ is the random variable, determining the index $\mu$ of the state 
 $\rho_\mu$, $K$ is the random variable for the measurement outcome $k$, and 
 $H(X)$ denotes the Shannon entropy of a random variable $X$.
However, $\hat d_\mathrm{ce} = 0$ already implies $\hat d_\mathrm{mf}= 0$ 
 since $d_\mathrm{ce}(\Xambda)\ge d_\mathrm{mf}(\Xambda)$ holds for any 
 measurement $\Xambda$ \footnote{%
In order to see this, note that for any probability distribution $P=(p_\mu)$ we 
 have $1-\max_\mu p_\mu \le -\log_b \max_\mu p_\mu \le H(P)$ for any base $1< 
 b\le \mathrm e$.%
}.

---
At this point the moderately impatient reader may directly skip to our main 
 result summarized in Proposition~\ref{p25379}.
Otherwise, allow us to introduce some additional notation:

First we combine the \emph{a priori} probabilities $p_\mu$ and the states 
 $\rho_\mu$ to \emph{weighted} states $\gamma_\mu\equiv p_\mu \rho_\mu$.
For a moment let us assume, that the measure $d$ is defined for arbitrary 
 families of $N$ weighted states with $\sum_\mu \tr\gamma_\mu = 1$ (this will 
 be guaranteed by property (ia) of regular measures we are about to define).
Then we write $d(\Xambda)\equiv d[\Xambda;(\gamma_\mu)]$ and let for an 
 operator $A$
\begin{equation}\label{e28263}
 d(\Xambda|A)= \begin{cases} d[\Xambda;(A\gamma_\mu A^\dag / p_A) ]) & \text{if 
 $p_A>0$}\\
 d[\idop;(\gamma_\mu)] & \text{else},\end{cases}
\end{equation}
 where $p_A= \sum_\mu \tr(A\gamma_\mu A^\dag)$ and $\idop\colon X \mapsto 
 \tr[X]$ is the trivial measurement.
(The operator $A$ in this definition shall correspond to the Kraus operator of 
 a measurement result, i.e.\ $A^\dag A$ is an effect of a POVM.
Then $d(\Xambda|A)$ denotes the deviation, given that we have performed a 
 certain POVM and obtained the result with effect $A^\dag A$.)

Although we will focus on the measure $d_\mathrm{mf}$, most parts of our 
 method apply to general regular deviation measures:
We call a deviation measure $d$ for $N$ states \emph{regular}, if the following 
 conditions are satisfied:
\begin{itemize}
\item[(ia)]
The measure $d[\Xambda;(\gamma_\mu)]$ only depends on $p(\mu,k)\equiv 
 \Xambda(\gamma_\mu)_k$; $d$ is well-defined for all probability distributions 
 $p(\mu,k)$ with $p(\mu,k)\ge 0$ and $\sum_{\mu,k} p(\mu,k)= 1$.
\item[(ib)]
For a fixed number of measurement outcomes, $d$ is bounded and continuous in 
 $p(\mu,k)$.
\item[(ii)]
A classical post-processing \footnote{%
A classical post-processing can be described by a stochastic matrix 
 $\pi_{\ell|k}$ with $\sum_\ell \pi_{\ell|k}=1$, such that 
 $\Pi[\Xambda(\rho)]_\ell= \sum_k \pi_{\ell|k}\Xambda(\rho)_k$.
}
$\Pi$ acts non-decreasing, i.e., $d(\Pi\circ \Xambda)\ge d(\Xambda)$.
\item[(iii)]
If a measurement is performed in two stages, then optimal post-selection after 
 the first stage acts non-increasing.
That is, if $\Xambda$ is of the form $\Xambda\colon \rho\mapsto \bigoplus_k 
 \Xambda_k(A_k \rho A_k^\dag)$ with $\sum_k A_k^\dag A_k= \id$ and measurements 
 $\Xambda_k$, then $d(\Xambda)\ge \min_k d(\Xambda_k|A_k)$.
\end{itemize}
We mention, that condition (iii) is satisfied for $d_\mathrm{mf}$ and 
 $d_\mathrm{ce}$ due to $d(\Xambda)= \sum_k p_{A_k} d(\Xambda_k|A_k)$ for 
 either measure; $d_\mathrm{mf}$ and $d_\mathrm{ce}$ in particular are 
 regular.
On the other hand, the measure $d_\mathrm{finite}$ satisfies all conditions but 
 the continuity condition in (ib).

\section{A necessary condition for perfect asymptotic discrimination}
\label{s20497}
In this section we will derive our main result, Proposition~\ref{p25379}, which 
 states a necessary condition for perfect discrimination by asymptotic LOCC, 
 $\hat d_\mathrm{mf}= 0$.
We present this proof in four steps:
As a prelude we will start with pseudo-weak measurements, a technique that will 
 become important for the protocol splitting method.
The protocol splitting (cf.\ Ref.~\cite{Bennett:1999PRA}) then achieves a split 
 of the protocol into stage~I and a continuation of stage~I.
This in turn allows to genuinely bound $\hat d$, cf.\ Eqns.~\eqref{e4911} and 
 \eqref{e10534}.
Finally we specialize this intermediate result to the regular deviation measure 
 $d_\mathrm{mf}$, yielding Proposition~\ref{p25379}.

\subsection{Prelude: Pseudo-weak measurements}\label{s13221}
Given a POVM $(E_k)$ we define for $b_k\ge 0$ and $\beta\equiv 1/(1+\sum_k 
b_k)$ the POVM $(E_k^\mathrm{pw})$ and the family of POVMs 
$(E_{(k),\ell}^\mathrm{rc})$ via
\begin{subequations}
\begin{align}
 E_k^\mathrm{pw}&=\beta\, (b_k \id+ E_k),\\
 E_{(k),\ell}^\mathrm{rc}&=\beta\, (b_k + \delta_{k,\ell})
   (E_k^\mathrm{pw})^{-1/2} E_\ell (E_k^\mathrm{pw})^{-1/2},
\end{align}
\end{subequations}
 with $\delta_{k,l}=1$ if $k=l$ and zero else --- if $b_k= 0$, we let 
 $E^\mathrm{rc}_{(k),l}= \delta_{k,l} \id$.
A measurement of $(E^\mathrm{pw}_k)$ is a \emph{pseudo-weak} implementation of 
 $(E_k)$, while we will refer to $(E^\mathrm{rc}_{(k),\ell})$ as the 
 \emph{recovery} measurement for outcome $k$.
Indeed, an application of the recovery measurement after the pseudo-weak 
 measurement on $\ket\psi$ results in
\begin{equation}
 U_{k,\ell} \sqrt{E^\mathrm{rc}_{(k),\ell}}\, \sqrt{E^\mathrm{pw}_k}\ket{\psi}
 = \sqrt{\beta\, (b_k+  \delta_{k,l})}\,\sqrt{E_\ell}\, \ket{\psi},
\end{equation}
 with $U_{k,\ell}$ a unitary originating from the polar decomposition.
In particular, if the outcome of the pseudo-weak measurement is ignored, the
 (weighted) state for outcome $\ell$ is identical to the state obtained by the 
 original measurement in the case of outcome $\ell$.

Let us now consider a completely positive and trace preserving (CPTP) map 
 $\Lambda$ described by Kraus operators $(A_k)$, $\Lambda\colon \rho\mapsto 
 \sum_k A_k \rho A_k^\dag$.
With $A_k= V_k\sqrt{E_k}$ a polar decomposition of $A_k$ (where $V_k^\dag V_k= 
 \id$), this map corresponds to a measurement of the POVM $(E_k)$ and a 
 subsequent application of $V_k$ and
 hence we can use the above method to obtain a pseudo-weak implementation 
 $(A_k^\mathrm{pw})$ of $(A_k)$ via $A_k^\mathrm{pw}=\sqrt{E_k^\mathrm{pw}}$.
The recovery step is then a CPTP map described by $(A_{(k),\ell}^\mathrm{rc})$ 
 with $A_{(k),\ell}^\mathrm{rc}=V_\ell 
 U_{k,\ell}\sqrt{E_{(k),\ell}^\mathrm{rc}}$.

\subsection{Protocol splitting}
\begin{figure}
\begin{center}
\includegraphics[width=.45\textwidth]{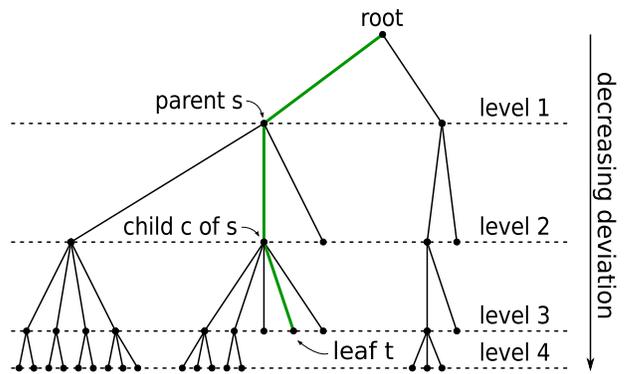}
\end{center}
\caption{\label{f19294}
Example of a {4}-leveled tree graph which represents an LOCC measurement with 
 at most {4} steps.
The branch $B(t)$ (thick green) connects the leaf $t$ with the root node and 
 hence consists of the root node, node $s$ and its child $c$ and the leaf $t$.
}
\end{figure}

In general, a finite LOCC protocol consists of a certain number of steps, where 
 in each step a particular party applies a family $(\Lambda_k)$ of local 
 quantum operations $\Lambda_k\colon \rho\mapsto A_k \rho A_k^\dag$ with 
 $\Lambda= \sum_k \Lambda_k$ trace preserving.
These quantum operations depend on the course of the protocol so far and the 
 measurement result $k$ is always communicated to all parties.
This situation can be depicted by a tree graph (cf.\ Fig.~\ref{f19294}), where 
 the children of each node correspond to a particular operation $\Lambda_k$, a 
 level in the tree represents a particular protocol step, and each branch 
 corresponds to a particular course of the protocol.

Hence, a finite LOCC protocol can be represented by a tree graph with root 
 element, where to each node $s$ of the tree, an operator $A_{(s)}$ is 
 associated.
(The associated operator for the root node is the identity operator.)
For each node, the associated child operators $(A_{(c)})$ shall form a family 
 of Kraus operators of a local CPTP map, i.e., all operators in $(A_{(c)})$ act 
 only non-trivially on some particular party.
Then for any path $P$ in this tree we associate an operator $A_P$ as the 
 product of the operators in reversed ordering:
If $P=(s_1,\dotsc,s_m)$, where $s_k$ is the parent of $s_{k+1}$, then 
 $A_P=A_{(s_m)}\cdots A_{(s_1)}$.
Note that $A_P$ is a product operator.
For a node $s$ we then denote by $B(s)$ the path connecting the root element 
 with $s$ (including the root element and $s$).

\begin{figure}
\begin{center}
\includegraphics[width=.45\textwidth]{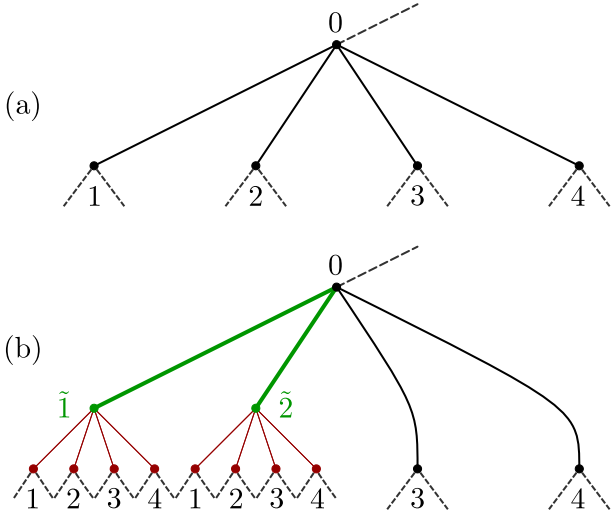}
\end{center}
\caption{\label{f16850}
Introduction of a pseudo-weak measurement and recovery step.
Assume that $d(\idop|A_{B(1)})\le  d(\idop|A_{B(2)}) < \delta$, while $\delta 
 \le d(\idop|A_{B(3)})\le  d(\idop|A_{B(4)})$ in the original situation (a).
In (b) the pseudo-weak measurement was introduced with $b_{(1)}$ and $b_{(2)}$ 
 such that $d(\idop|A_{B(\tilde 1)})= d(\idop|A_{B(\tilde 2)})= \delta$ (thick 
 green), while the operators at nodes $3$ and $4$ remain ---up to a 
 prefactor--- unchanged.
Then for the nodes $\tilde 1$ and $\tilde 2$ a recovery step is introduced 
 (thin red), such that ---up to a prefactor--- in effect the original operators 
 from the nodes $1$ to $4$ occur.
Finally the according parts of the original protocol are added to the outcomes 
 of the recovery measurements (dashed gray).
}
\end{figure}
For an arbitrary $\delta$ with $0<\delta< d(\idop)$ (again, $\idop\colon 
 X\mapsto \tr[X]$) we modify the protocol in an iterative procedure as follows 
 (cf.\ Fig.~\ref{f16850}).
For any node $s$ we denote by $D_\delta(s)$ the set of child nodes for which 
 the deviation dropped below $\delta$, i.e.,
\begin{equation}
 D_\delta(s) = \{\text{$c$ is child of $s$} \mid d(\idop|A_{B(c)}) < \delta \}.
\end{equation}
Let $s$ be a node with non-empty set $D_\delta(s)$ but $d(\idop|A_{B(a)})> 
 \delta$ for any $a\in B(s)$.
For such a node, the associated child operators $(A_{(c)})$ are replaced by the 
 pseudo-weak implementation $(A^\mathrm{pw}_{(c)})$ with the parameters 
 $(b_{(c)})$ (cf.\ Sec.~\ref{s13221}) chosen such that 
 $d(\idop|A^\mathrm{pw}_{(c)} A_{B(s)})= \delta$ for all $c\in D_\delta(s)$ and 
 $b_{(c)}= 0$ else.
This is always possible, since regular deviation measures are continuous and 
 the pseudo-weak measurement smoothly interpolates between $A_{(c)}\equiv 
 V_{(c)} \sqrt{A^\dag_{(c)}A_{(c)}}$ and $V_{(c)}$ for $b_{(c)}= 0\dotsc 
 \infty$.
For the nodes in $D_\delta(s)$ we add the recovery step as an additional level 
 (the recovery measurement for the remaining child nodes would be trivial).
After the recovery measurement, the according part of the original protocol is 
 appended.

This procedure is repeated, until for all nodes $s$ either $D_\delta(s)$ is 
 empty or there exists an $a\in B(s)$ with $d(\idop|A_{B(a)})= \delta$.
It is important to note, that this procedure terminates after a finite number 
 of steps.
This is the case, since the number of candidates subject to modification 
 decreases in each step of the procedure; the recovery levels are only 
 introduced when $d(\idop|A_{B(s)})= \delta$.

We denote by \emph{stage~I} of the protocol the part that does not enter the 
 recovery steps, but rather terminates as soon as $d(\idop|A_{B(s)})= \delta$ 
 in the modified protocol.

\subsection{Analysis of the best-case deviation}
For the moment we only consider stage~I of the modified protocol (with 
 parameter $\delta$).
As an abbreviation we define for each leaf $k$ of this stage the shorthand 
 $A_k:= A_{B(k)}$.
Let us now define the set
\begin{equation}
 S_\delta= \{\text{$k$ is leaf} \mid d(\idop|A_k)= \delta \}.
\end{equation}
Due to our modification of the protocol, $k\notin S_\delta$ only if $k$ was 
 already a leaf in the original protocol with $d(\idop|A_k)> \delta$.

For each leaf $k$ we let $\Xambda_k$ be the continuation of stage~I of the 
 modified protocol.
With $\Pi$ being the post-processing that ``forgets'' all results of any 
 pseudo-weak measurement introduced by the protocol splitting (this are those 
 results with parameter $b_{(c)}>0$), the measurement
\begin{equation} \textstyle
 \Xambda\colon \rho\mapsto \Pi[ \bigoplus_k \Xambda_k(A_k\rho A_k^\dag) ]
\end{equation}
 is equivalent to the original protocol.
Hence, due to property (ii) and (iii) of regular measures $d$, we have
\begin{equation}
 d(\Xambda) \ge \min_k d(\Xambda_k|A_k)
 \ge \min[\delta, \min_{k\in S_\delta} d(\Xambda_k|A_k)].
\end{equation}

We now consider the case of $\hat d= 0$, i.e., for any $\varepsilon>0$ there 
 exists a protocol $\Xambda^\varepsilon$ with 
 $d(\Xambda^\varepsilon)<\varepsilon$.
Then for any $\delta$ with $0< \delta< d(\idop)$ and any $\varepsilon$ with $0< 
 \varepsilon< \delta$ we have
\begin{equation}
 \varepsilon > d(\Xambda^\varepsilon)
 \ge\min_{k\in S_\delta} d(\Xambda_k|A_k).
\end{equation}
(Note that $\Xambda_k$ and $A_k$ depend on $\delta$ and $\Xambda^\varepsilon$.)
The right-hand side of this inequality can be further lower bounded by
\begin{subequations}\label{e4911}
\begin{multline}\label{e4911a}
 y_\delta= \inf\{d(\mathfrak G|\sqrt E) \mid
  \text{$\mathfrak G$ is a finite LOCC}\\\text{measurement, } E\in M_\delta\},
\end{multline}
 where
\begin{multline}\label{e4911b}
 \mathcal M_\delta= \{ E\text{ is a product operator}\mid E\ge 0,\\
  {\textstyle \sum_\mu} \tr(\gamma_\mu E)= 1 \text{, and }
  d(\idop|\sqrt E)= \delta\}.
\end{multline}
\end{subequations}
This is a lower bound, since any $\Xambda_k$ is a finite LOCC measurement, 
 $A_k^\dag A_k/p_{A_k}\in \mathcal M_\delta$ [cf.\ Eq.~\eqref{e28263}; the case 
 $p_{A_k}= 0$ cannot occur due to $\delta<d(\idop)$], and due to property (ia) 
 of regular deviation measures.
We have an intermediate result:
\begin{equation}\label{e10534}\text{
$\hat d= 0$ only if $y_\delta= 0$ for any $0< \delta< d(\idop)$.
}\end{equation}

The main use of this result is the reverse statement, where $y_\delta> 0$ for 
 some $\delta$ shows that $\hat d>0$.
In this case we are not interested in the actual value of $y_\delta$, and we 
 therefore now aim to eliminate the infimum in the expression for $y_\delta$.

\subsection{Specialization to $d_\mathrm{mf}$}
The special property of the measure $d_\mathrm{mf}$, as defined in 
 Eq.~\eqref{e3208}, we are about to exploit is, that for the discrimination of 
 $N$ states, it is never advantageous to choose a measurement with more than 
 $N$ outcomes (for more than $N$ outcomes one could always combine the results 
 for which $\max_\nu p_\nu \mathfrak E(\rho_\nu)_k$ is achieved at $\mu=\nu$).
Therefore, in order to make the set of measurements in the definition of 
 $y_\delta$ [cf.\ Eq.~\eqref{e4911a}] a compact set, we extend the allowed 
 measurements to arbitrary global measurements \footnote{%
Using global measurements yields a rather rough estimate.
One could also choose the set of fully separable measurements as defined in 
 Eq.~\eqref{e19171}.
Then Proposition~\ref{p25379} would contain the additional restriction, that 
 the states must allow a discrimination by separable measurements after result 
 $E$.
This condition, however, is difficult to evaluate and hence we only cover the 
 simpler case of global measurements here.
}, but at the same time consider only measurements with at most $N$ outcomes.

We also assume that the kernels of the states $(\gamma_\mu)$ do not share a 
 product vector, i.e., $\bigcap_\mu\ker\gamma_\mu$ contains no product vector 
 (except $0$).
Let $E\in \mathcal M_\delta$, as defined in Eq.~\eqref{e4911b}, have the 
 spectral decomposition $E=\sum_j e_j \proj{j}$, where $\ket{j}$ are product 
 vectors.
Then with $R=\sum_\mu \gamma_\mu$ we have
\begin{equation}
 1= \tr(RE) \ge \min_j( \bra j R \ket j)\max_j e_j \ge  \eta_R\max_j e_j.
\end{equation}
 where $\eta_R= \inf \bra\xi R \ket\xi$, with the infimum taken over \emph{all} 
 product vectors $\ket\xi$.
Since the kernel of $R$ contains no product vector, $\eta_R>0$ and hence 
 $e_j\le 1/\eta_R$.
This in turn shows that we can replace the condition $E\ge 0$ by the compact 
 condition $\id/\eta_R \ge E \ge 0$.
Due to the condition $\sum_\mu \tr \gamma_\mu E= 1$, we have $d(\idop|\sqrt 
 E)=d[\idop; (\sqrt E \gamma_\mu \sqrt E)]$ which shows due to the continuity 
 of $d$, that the condition $d(\idop|\sqrt E)= \delta$ defines a compact set.
Hence $\mathcal M_\delta$ as defined in Eq.~\eqref{e4911b} itself is a compact 
set.

Together with the continuity of regular measures, it follows that $\hat 
 d_\mathrm{mf}=0$ only if there exists an operator $E$ in $\mathcal M_\delta$ 
 and a measurement $\mathfrak G$ with $d(\mathfrak G|\sqrt E)= 0$.
Hence the states $(\sqrt E \gamma_\mu \sqrt E)$ can be perfectly discriminated 
 and thus are mutually orthogonal, i.e.\ $\tr(\gamma_\mu E \gamma_\nu E)= 0$ 
 for $\mu\ne \nu$.

Finally, our argument is independent of the \emph{a priori} probabilities 
 $p_\mu>0$, and we hence can choose them to be all equal (this maximizes 
 $d_\mathrm{mf}(\idop)$ to $1/N$ and hence the range of $\delta$).
The boundary cases $\delta= 0$ and $\delta= d_\mathrm{mf}(\idop)$ are trivial 
 to fulfill.
Letting $\chi=1-\delta$, we arrive at our main result:
\begin{proposition}\label{p25379}
Let $(\rho_\mu)$ be a family of $N$ states, such that $\bigcap_\mu 
 \ker\rho_\mu$ contains no product vector (except 0).

Then $(\rho_\mu)$ can be discriminated perfectly by asymptotic LOCC, $\hat 
 d_\mathrm{mf}= 0$, only if for all $\chi$ with $1/N \le \chi \le 1$ there 
 exists a product operator $E\ge 0$ obeying $\sum_\mu \tr(E \rho_\mu)= 1$,
 $\max_\mu \tr(E \rho_\mu)= \chi$, and $\tr(E \rho_\mu E\rho_\nu)= 0$ for 
 $\mu\ne \nu$.
\end{proposition}

This necessary condition does not imply perfect discrimination for finite 
 LOCC, as we will demonstrate in Section~\ref{s2614}.
We mention, that the Proposition basically holds for any regular deviation 
 measure $d$, for which the optimal general measurement strategy for $N$ 
 arbitrary states can be achieved using at most a certain fixed number of 
 effects.

Note, that the precondition in Proposition~\ref{p25379} is not robust under 
 trivial local embeddings:
If a local Hilbert space $\Hilbert^{(s)}$ is extended to $\Hilbert^{(s)}\oplus 
 \Hilbert'$, this condition will be violated.
However, if $E'\in \mathcal M_\delta$, then the projection onto the original 
 space $E'\mapsto E$ is still in $\mathcal M_\delta$ and $d(\mathfrak G|\sqrt 
 E)= d(\mathfrak G|\sqrt{E'})$.
Therefore, in the Proposition the embedding Hilbert space $\Hilbert= 
 \bigotimes_r \Hilbert^{(r)}$ should be chosen as small as possible.

\section{Examples}\label{s17873}
\subsection{Product bases}
Let $(\ket{\psi_\mu})$ be an orthonormal product basis of an $N$-dimensional 
 Hilbert space $\Hilbert= \bigotimes_r \Hilbert^{(r)}$.
We assume that the states $(\proj{\psi_\mu})$ can be discriminated by 
 asymptotic LOCC and hence for any $\chi$ with $1/N\le \chi\le 1$ there exists 
 an operator $E$ obeying the conditions in Proposition~\ref{p25379}.
For $1/N< \chi< 1/(N-1)$, this operator $E$ must be of full rank, but cannot be 
 a multiple of the identity operator.

We choose some decompositions $\ket{\psi_\mu}= \bigotimes_r 
 \ket{\omega_\mu^{(r)}}$ and $E=\bigotimes_r E^{(r)}$ with $E^{(r)}\ge 0$.
Since $\bra{\psi_\nu}E\ket{\psi_\mu}= 0$ if and only if $\mu\ne \nu$, it 
 follows that $E\ket{\psi_\mu}= f_\mu \ket{\psi_\mu}$ with $f_\mu >0$.
Hence for any $r$ we have $E^{(r)} \ket{\omega_\mu^{(r)}}= f_\mu^{(r)} 
 \ket{\omega_\mu^{(r)}}$ with $f_\mu^{(r)}>0$.
It follows that a local measurement of the \emph{observable} $E^{(r)}$ does not 
 change any of the input states.
Since for some subsystem $s$, the observable $E^{(s)}$ is not proportional to 
 the identity operator, the measurement of $E^{(s)}$ separates the set of 
 states in at least two non-empty subsets.
Each of the subsets is again an orthonormal product basis of a subspace of 
 $\Hilbert$ and each of the subsets inherits the property that it can be 
 discriminated by asymptotic LOCC.
By induction we arrive at
\begin{proposition}\label{p23315}
If a complete (product) basis can be discriminated perfectly by asymptotic LOCC 
 ($\hat d_\mathrm{mf}= 0$) then it can already be discriminated perfectly by a 
 finite LOCC measurement.
\end{proposition}

Since $\hat d_\mathrm{mf}> 0$ implies $\hat d_\mathrm{ce}> 0$ [cf.\ 
 Eqns.~\eqref{e3208} and \eqref{e21158}], this Proposition in particular yields 
 an analytical proof of the result of Bennett \emph{et al.} in 
 Ref.~\cite{Bennett:1999PRA}.
Unfortunately, it is not straightforward to extend this type of argument to the 
 situation of an unextendible product basis (then $\ket{\psi_\mu}$ is not 
 necessarily an eigenstate of $E$.).

\textit{Remark.}
In Ref.~\cite{Rinaldis:2004PRA} a proof was given that unextendible product 
 bases cannot be discriminated by asymptotic LOCC.
(Since a complete basis is also unextendible, this includes 
 Proposition~\ref{p23315} as a special case.)
While the statement is likely to hold, the proof given there is incomplete.
In particular we question the argument below Eq.~(16), showing that the 
 quotient ``$M_N/c_N$'' converges to a constant for ``$N \rightarrow\infty$'' 
 (in this expression $N$ denotes the number of steps until the protocol is 
 aborted).
The argument for this convergence is quite general and should hold whenever 
 finite discrimination is not possible (more precisely, if any local 
 measurement either destroys orthogonality or is trivial).
For the example in Sec.~\ref{s2614}, however, the quotient would diverge, since
 ``$c_N$'' is zero in this case.

\subsection{When Proposition~\ref{p25379} does not decide}\label{s2614}
The previous example showed that for a wide class of examples, asymptotic LOCC 
 does not provide an advantage over LOCC with finite resources.
In this section we give an explicit example for which Proposition~\ref{p25379} 
 does not help to decide whether perfect discrimination via asymptotic LOCC can 
 be performed.

We aim to discriminate the following three mutually orthogonal states on a 
two-qubit system:
\begin{equation}\label{e8361}\begin{split}
 \ket{\psi_1}&=\ket{00},\\
 \ket{\psi_2}&\propto
   2\ket{01}-(\sqrt 3+1)\ket{10}-\sqrt{6}\sqrt[4]{3} \ket{11},\\
 \ket{\psi_3}&\propto
   2 \ket{01} - (\sqrt 3-1) \ket{10}+ \sqrt{2} \sqrt[4]{3} \ket{11}.
\end{split}\end{equation}
In Ref.~\cite{Duan:2009IIT}, Example~1 \footnote{We chose $\alpha=\pi/12$, 
 $\beta=\pi/6$, and $\tan\gamma=3^{-1/4}$ and different local bases.}, it has 
 been demonstrated, that this set of vectors can be discriminated perfectly by 
 stochastic LOCC, while there exists no perfect discrimination strategy for 
 LOCC in a finite number of steps.
In fact, a local effect that does not destroy orthogonality is necessarily 
 proportional to the identity operator.

The only state that is orthogonal to all $\ket{\psi_\mu}$ is entangled and 
 hence we can apply Proposition~\ref{p25379}.
However, in the Appendix we construct an operator $E_\chi$ for $\frac 13 \le 
 \chi \le 1$, which satisfies the conditions from Proposition~\ref{p25379}.
Hence our necessary condition for perfect discrimination by asymptotic LOCC is 
 satisfied, but Proposition~\ref{p25379} does not provide a sufficient 
 criterion.

\section{Conclusions}\label{s25406}
We considered the case of asymptotic local operations and classical 
 communication for the discrimination of mutually orthogonal states and 
 derived a necessary condition for perfect asymptotic discrimination to be 
 possible.
Our analysis yielded a general necessary condition, cf.\ 
 Proposition~\ref{p25379}, which consists of the existence of a certain product 
 operator.
As an example we showed, that any complete basis of product states can be 
 discriminated perfectly by asymptotic LOCC if and only if they can already be 
 discriminated in a finite number of rounds (cf.\ Proposition~\ref{p23315}).

Our result allows to relatively easily exclude whether a family of states can 
 be discriminated by asymptotic LOCC, however it is still unclear whether 
 infinite resources can be of any advantage.
Although the general intuition might be, that for perfect discrimination the 
 asymptotic case is not superior, we identified an example, which could be a 
 counter-example for this case as our necessary condition is fulfilled.
However, as a sufficient criterion is not available, this question remains 
 open.

\begin{acknowledgments}
We thank
O.~Gittsovich,
O.~G\"uhne,
B.~Jungnitsch,
B.~Kraus,
T.~Moroder,
S.~Niekamp, and
A.~Thiel
for helpful discussions.
This work has been supported by the DFG and
the Austrian Science Fund (FWF): Y376 N16 (START Prize) and SFB FOQUS.
\end{acknowledgments}

\appendix*

\section{Construction of $E_\chi$}
In this Appendix we provide an operator $E_\chi$ for the states defined in 
 Eq.~\eqref{e8361}.
This operator satisfies the conditions from Proposition~\ref{p25379}.
We first define the local qubit-operator $A_\chi$ via
\begin{equation}\begin{split}
 \bra{0}A_\chi\ket{0}&=
  -(12 \sqrt{3}-21) \chi +3\sqrt{3}-3,\\
 \bra{1}A_\chi\ket{1}&= (6\sqrt{3}-12) \chi -2 \sqrt{3}+6,\\
 \bra{0}A_\chi\ket{1}&=\textstyle
  \sqrt{2 \sqrt{3}-3} [(5 \sqrt{3}-3) \chi -2 \sqrt{3}],\\
 \bra{1}A_\chi\ket{0}&= \bra{0}A_\chi\ket{1}^*,
\end{split}\end{equation}
 and the diagonal operators $B_\chi$ and $C_\chi$ via
\begin{equation}\begin{split}
\bra{0}B_\chi\ket{0}&=20 \chi+2 \tilde\chi-4,\\
\bra{1}B_\chi\ket{1}&=(12-\sqrt3)\chi + \tilde\chi+\sqrt3-1,
\end{split}\end{equation}
 and
\begin{equation}\begin{split}
\bra{0}C_\chi\ket{0}&=-(4+3\sqrt3)\chi-\tilde\chi+3\sqrt3+5,\\
\bra{1}C_\chi\ket{1}&= \bra1B_\chi\ket1,
\end{split}\end{equation}
 where
\begin{equation}
 \tilde\chi=\sqrt{(115-8\sqrt 3)\chi^2-(46-10\sqrt3)\chi-2\sqrt3+4}.
\end{equation}

Then with
\begin{equation}
 \tilde E_\chi=\begin{cases}
   B_\chi\otimes C_\chi & \text{if $\chi<1/2$}\\
   A_\chi\otimes \proj 1 & \text{else.}
 \end{cases}
\end{equation}
 we finally let $E_\chi= \tilde E_\chi / \sum_\mu \bra{\psi_\mu}\tilde E_\chi 
 \ket{\psi_\mu}$.
One readily verifies that $E_\chi$ has the desired properties.

\bibliography{the}

\end{document}